\documentclass{aa}  
\usepackage{times}
\usepackage[T1]{fontenc}
\usepackage{pslatex}
\usepackage{amsmath}
\usepackage{color,graphicx}
\usepackage{rotating}
\usepackage{txfonts}
\begin{document} 
\title{RAVE stars tidally stripped/ejected from $\omega$ Centauri globular cluster} 
\author{
 J. G. Fern\'andez-Trincado\inst{1}, 
 A. C. Robin\inst{1},
 K. Vieira\inst{2},
 E. Moreno\inst{3},
 O. Bienaym\'e\inst{4},
 C. Reyl\'e\inst{1},
 O. Valenzuela\inst{3},
 B. Pichardo\inst{3},
 F. Robles-Valdez\inst{5},
 \and
 A. M. M. Martins\inst{1}
 }

 \institute{Universit\'e de Franche-Comt\'e, Institut Utinam, UMR CNRS 6213, OSU Theta, Besan\c{c}on, F-25010, France.\\
 \email{jfernandez@obs-besancon.fr}
 \and
 Centro de Investigaciones de Astronom\'ia, AP 264, M\'erida 5101-A, Venezuela.
 \and
 Instituto de Astronom\'ia, Universidad Nacional Aut\'onoma de M\'exico, Apdo. Postal 70264, M\'exico D.F., 04510, Mexico.
 \and
 Observatoire astronomique de Strasbourg, Universite de Strasbourg, CNRS, UMR 7550, 11 rue de l'Universit\'e, F-67000 Strasbourg, France.
 \and
  Instituto de Ciencias Nucleares, Universidad Nacional Autónoma de México, Ap. 70-543, 04510, D.F., México, Mexico.
 }

 \date{Received 26/05/2015; Accepted 10/08/2015}
 \titlerunning{$\omega$ Centauri}

\authorrunning{J. G. Fern\'andez-Trincado et al.} 

 
  \abstract
   {Using six-dimesional phase-space information from the Fourth Data release of the Radial Velocity Experiment (RAVE)
over the range of Galactic longitude 240$^{\circ}< l <$ 360$^{\circ}$ and $V_{LSR} < -239$ kms$^{-1}$, 
we have computed orbits for 329 RAVE stars 
that were originally selected as chemically and kinematically related to $\omega$ Centauri. 
 The orbits were integrated in a Milky-Way-like axisymmetric Galactic potential,
ignoring the effects of the dynamical evolution of $\omega$ Centauri due to the tidal effects of the Galaxy disk on the cluster along time. We also ignored secular changes in the Milky Way potential over time.  
In a Monte Carlo scheme, and under the assumption that the stars may have been ejected with velocities greater than the escape velocity ($V_{rel}>V_{esc,0}$) from the cluster, we identified 15 stars as having close encounters with $\omega$ Centauri: (\textit{i}) 8 stars with relative velocities $V_{rel}< 200 $ kms$^{-1}$  may have been ejected $\sim$ 200 Myr ago from $\omega$ Centauri; (\textit{ii}) other group of 7 stars were identified with high relative velocity $V_{rel}> 200 $ kms$^{-1}$ during close encounters, and seems unlikely that they have been ejected from $\omega$ Centauri. We also confirm the link between J131340.4-484714 as potential member of $\omega$ Centauri, and probably ejected $\sim$ 2.0 Myr ago, with a relative velocity $V_{rel}\sim80$ kms$^{-1}$. }     
   \keywords{Stars: abundances - Stars: kinematics and dynamics - Galaxy: structure - Globular clusters: individual ($\omega$ Centauri, NGC 5139) }
   \maketitle
\section{Introduction}
\label{Section1:Introduction}

Known as the most massive and luminous globular cluster of the Milky Way's Halo, 
$\omega$ Centauri also shows unique physical properties in its structure, chemical enrichment and internal kinematics
\citep{Freeman1975, Norris1997, Lee+1999, Hughes+2000, Gnedin+2002, Mizutani2003, Pancino+2003, Hilker+2004, Bedin+2004, 
	Sollima+2007, Sollima2009, Bellini+2009, Johnson+2010, Bellini+2010, TakashiMoriya+2010, DaCosta2012}. 
The nature and origin of $\omega$ Centauri is still controversial.
It has been proposed that $\omega$ Centauri is the remnant core of a tidally disrupted satellite galaxy, 
possibly destroyed long ago by the Milky Way \citep{Lee+1999, Majewski2000, Bekki+2003, 
Mizutani+2003, Ideta+2004, Tsuchiya+2004, Romano2007, Sollima+2009}, and
some recent works suggest that it could have originated from 
an even more massive system of about $10^9$ M$_{\odot}$ \citep{Valcarce+2011}.\\

Under this scenario it is expected that the orbit of such progenitor would have evolved along time
because of the interaction with the Milky Way \citep{Tsuchiya2003, Bekki+2003, Zhao2004, Ideta+2004} and part of its residual material would be distributed in the Galactic Halo.
Several efforts have been made, with various degrees of success, to detect the relics of $\omega$ Centauri's progenitor. 
Among those that have explored the surroundings of the cluster, 
looking for the controversial pair of tidal tails extending from it and reported by \citet{Leon+2000}, 
are the effects of differential reddening as spurious effects on the detection of tidal tails \citep{Law2003}, and the spectroscopy survey of 4000 stars selected from the cluster red giant branch done by \citet{DaCosta+2008}, 
and the more recent photometric survey of RR Lyrae stars over $\sim$ 50 square degrees around the cluster 
by \citet[][]{Fernandez-Trincado2013, Fernandez-Trincado2015}. 
None of these works found evidence of stellar overdensities beyond the tidal radius of $\omega$ Centauri. 
However, different results are inferred from the photometric analysis of the STREGA survey \citep[STRructure and Evolution of the GAlaxy]{Marconi+2014},
whose tracers are mainly variable stars (RR Lyraes and long-period variables) and main-sequence turn-off stars.
These authors found evidence of stellar overdensities at $\sim 1^\circ$ from the cluster in the direction perpendicular 
to what was explored in the latter mentioned survey,
nonetheless a more detailed chemical abundances and kinematics analysis would help to verify this result.\\

Stellar debris features associated with $\omega$ Centauri have also been identified recently in a wide range of Galactic latitude, 
and more specially in the solar neighborhood, with chemical patterns and kinematics properties similar to inner populations of the cluster 
\citep{Nissen+2010, Wylie-deBoer+2010, Majewski+2012}, 
and more recently a group of RAVE stars with radial velocities and chemical abundances consistent with stars observed from $\omega$ Centauri cluster 
has been detected by \citep{Kunder+2014}. 
As mentioned by \citet{Meza+2005}, the presence of nearby stars with odd kinematics and chemistry, suggest an extra-Galactic origin for them, possibly ``tidal relicts''.\\
 From that starting point, we decided to investigate the kinematical properties of stars having preferential retrograde motion ($V_{LSR}< -239$ kms$^{-1}$), looking for the presence of $\omega$ Centauri debris.
 We adopted an idea similar to that of \citet[][]{Pichardo+2012} and combined 
the accurate radial velocities, distances and [Fe/H] abundances from the Fourth Data Release of the Radial Velocity Experiment survey 
(RAVE DR4) by \citet{Kordopatis+2013},  with proper motions from the US Naval Observatory CCD Astrograph Catalog 
(UCAC4) by \citet{Zacharias+2013}, to select a sample of giant stars that are chemically and kinematically similar to $\omega$ Centauri.
For each of star in the sample, we integrate $10^5$ orbits in a Milky-Way-like axisymmetric potential, 
to evaluate the probability of the occurrence of close previous encounters between the star and $\omega$ Centauri, within the cluster's tidal radius.\\

This paper is organized as follows: In \S\ref{selection} we describe the data and methods for the selection of star candidates. 
In \S\ref{kinematicontamination} we briefly explore the effects of contamination in our sample by different Galactic components. 
In \S\ref{experiment} we describe the orbits model. 
Results and discussion are presented in \S\ref{discusion}. 
Finally, conclusions are presented in \S\ref{conclusion}.\\

\section{Sample selection from RAVE}
\label{selection}

The sample was selected from the RAVE DR4 catalog \citep{Kordopatis+2013}, which provides accurate radial velocities with typical errors 
$\sigma_{RV}\sim$ 2 km s$^{-1}$, and distances and individual abundances with errors of about 10-20\%, 
determined for approximately 390,000 relatively bright stars (9 mag $< I_{DENIS} <$ 13 mag). 
The proper motions on RAVE DR4 were compiled from several catalogs, however, in this work we use UCAC4 \citep{Zacharias+2013}. 
We used these data to make a kinematical selection of RAVE stars possibly related to $\omega$ Centauri,
taking also into account spatial distribution and metallicity, as well as some additional quality control cuts to select robust data.\\ 

In this work we restricted our study to RAVE stars with Galactic longitudes $240^{\circ}<$l$<360^{\circ}$, 
where  $\omega$ Centauri's remnant candidates have been found \citep[e.g.,][]{Majewski+2012}.
We restricted our sample to giant stars, 
with effective temperature between 4000-5500 K, and surface gravity 0.5 dex $< log (g) <$ 3.5 dex,  following \citet{Boeche+2011}. 
Additionally, we required the stars to have high quality spectra ($\chi^2 < 2000$) with a signal-to-noise ratio S/N$>$20 
\citep[$algo\_conv=0$ was required, indicating that the pipeline converges, see ][]{Kordopatis+2013}. 
This cut allows us to obtain precise radial velocity measurements, typically $\sigma_{RV} < $ 2 kms$^{-1}$, in order to constraint the full space motion.
The metallicity [Fe/H] distribution for giant stars within $\omega$ Centauri spans more than a magnitude order, 
from -2.2 dex $ < $ [Fe/H] $< -0.7$  dex \citep{Johnson+2010}, therefore we allow stars in our sample to be in this range of metallicity.\\

A total of 9,024 RAVE stars in our final sample passed the selection and quality controls described above. 
Distances, radial velocity and proper motions are all available for this sample.\\

\begin{figure}
	\begin{center}
		\includegraphics[width=0.55\textwidth]{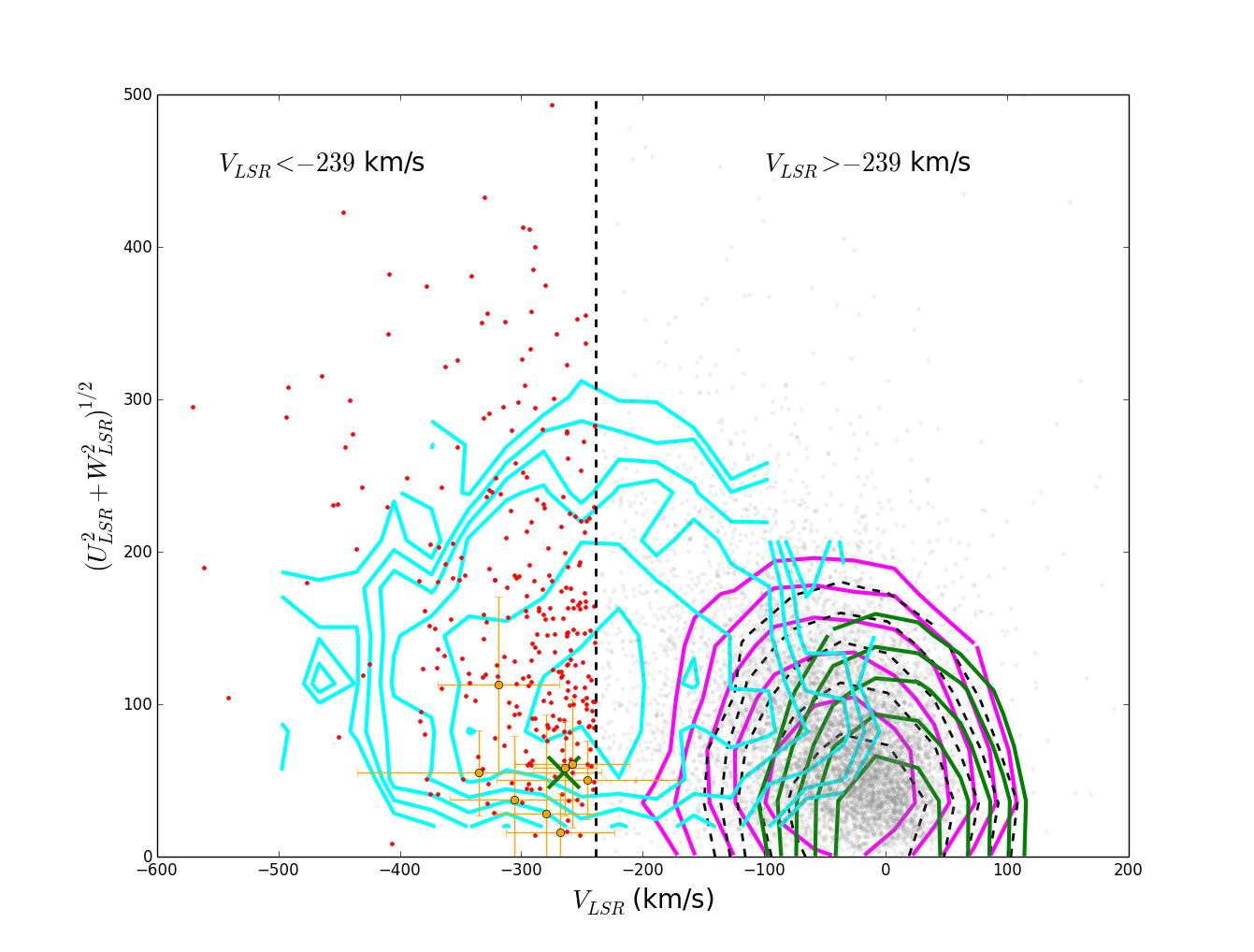}
	\end{center}
	\caption{Toomre diagram for RAVE stars and their simulation with the Besan\c{c}on Galaxy model. RAVE stars with quality control cuts (see \S\ref{selection}) are shown in grey dots. $\omega$ Centauri is shown with the green symbol.
		Isocontours refers to Thin-Disk (green), Young Thick-Disk (black dashed), Old Thick-Disk (magenta) and Halo (cyan) simulated RAVE stars.  
		Orange symbols (8 stars) with error bars correspond to RAVE stars having a high probability of encountering with $\omega$ Centauri and relative velocity $V_{rel}<200$ kms$^{-1}$ (see \S\ref{discusion}), and red dots (314 stars) correspond to Halo stars having a negligible probability ($< 0.01\%$) of encountering with $\omega$ Centauri (see \S \ref{experiment}). 
		The vertical dashed line indicates zero Galactic rotation. (A color version of this figure is available in the online journal.)} 
	\label{Toomre}
\end{figure}

\subsection{Kinematic selection}

With the RAVE kinematics and distances as input from the selected sample, 
galactic space velocities $(U,V,W)$ were computed according to the matrix equations of \citet{Johnson+1987}. 
For reference, we have adopted a right-handed coordinate system for $(U,V,W)$, 
so that they are positive in the directions of the galactic center, galactic rotation, and north galactic pole, respectively. 
In this convention, we have adopted the solar motion with respect to the Local Standard of Rest 
$(U_{\odot},V_{\odot},W_{\odot})=(11.10, 12.24, 7.24)$ km s$^{-1}$, 
assumed that the Local Standard of Rest is on a circular orbit with a circular speed of $239$ km s$^{-1}$, 
and the Sun is located at $R_{\odot}=8.3$ kpc \citep[e.g.,][]{Brunthaler+2011}. \\

Figure \ref{Toomre} shows the Toomre diagram for the Thin Disk, Young/Old Thick-Disk and Halo stars
from the revised version of the Besan\c{c}on Galaxy model \citep[][]{Robin+2014}. 
The grey dots represent the 9,024 RAVE stars previously selected. 
As seen in the Toomre diagram, most RAVE stars are kinematically consistent with the different Galactic components. 
Then, stars with $V_{LSR} < -239$ kms$^{-1}$ are more probably Galactic halo stars, 
and this work is focused on studying the orbit of these stars and their possible connection with $\omega$ Centauri.\\ 

\citet{Majewski+2012} suggest that $\omega$ Centauri tidal debris are a primary contributor of retrograde velocity stars near the Sun. 
In fact, evidence has been found of field stars associated with $\omega$ Centauri in the Galactic Disk by \citet[][]{Wylie-deBoer+2010}.
The cut $V_{LSR}<-239$ kms$^{-1}$ acts as kinematical membership selection and helps us to obtain 
a sample of 329 stars that is mostly free of contamination by the Galaxy's Thin and Thick Disks.
Given the magnitudes of these stars we are able to explore past dynamic links
between these RAVE stars and $\omega$ Centauri.
For these stars we computed orbits using an axisymmetric model of the Galaxy scaling the model of \citet{Allen+1991}
(gravitational potential details can be seen at \S \ref{experiment}).
 \begin{figure*}
 	\begin{center}
 		\includegraphics[width=0.97\textwidth]{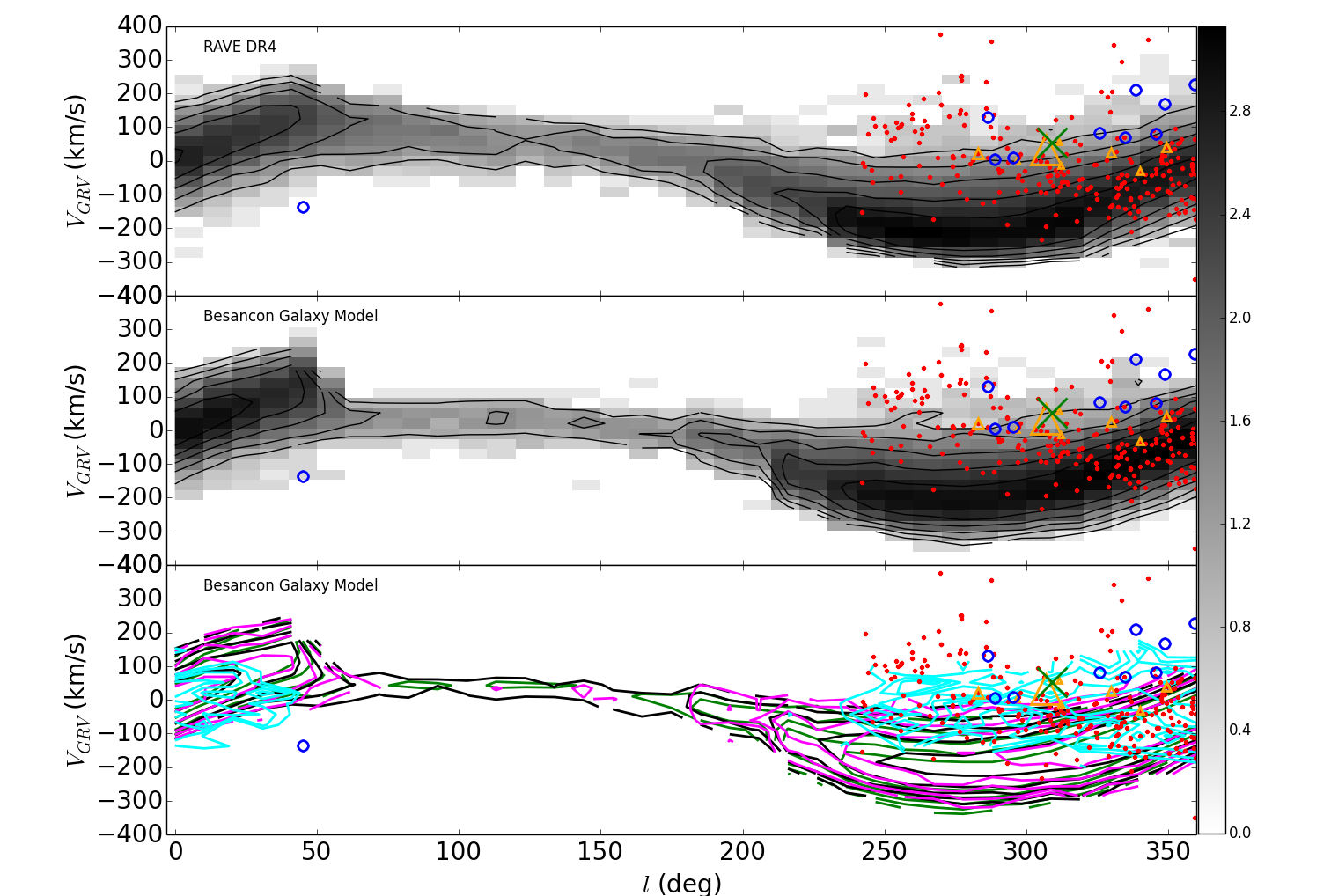}
 	\end{center}
 	\caption{RAVE stars in the Galactocentric radial velocity $(V_{GRV})$ vs. Galactic longitude space. 
 		Top panel refers to RAVE data, middle and bottom panel refers to Besan\c{c}on Galaxy model-simulated RAVE data. 
 		$\omega$ Centauri is shown with the green cross.
 		\citet{Majewski+2012} stellar "streams" are plotted in blue symbols.
 		Probable members found in this work are represented in orange symbols and are the same as in Figure \ref{Toomre}, their size is proportional to the probability of close encounter with the cluster (see Sect. \ref{experiment}), 
 		and RAVE Halo stars having a negligible probability of encountering with $\omega$ Centauri are shown with red dots.
 		The black contours show the full RAVE sample and simulations, respectively. 
 		The coloured contours refers to the Galactic components from the simulation as in Figure \ref{Toomre}.
 		The grey scale refers to the logarithm of the number of stars in it, as indicated by the colour bar. (A color version of this figure is available in the online journal.)} 
 \label{Density}
\end{figure*}

\section{Kinematic contaminants from the Milky Way}
\label{kinematicontamination}

In order to estimate the kinematical contamination in our sample by stars from the Galaxy, 
we simulated the RAVE data with the Besan\c{c}on Galaxy Model. 
We used the new version described in \cite{Robin+2014} where the Thick-Disk and Halo characterizations 
have been revised according to SDSS and 2MASS data. 
Moreover, the kinematics of the different populations is computed using a corrected asymmetric drift formula (Robin et al. 2015, in prep.) reliable out of the Galactic plane. 
The velocity ellipsoid of the Halo is assumed to be (131, 106, 85) km s$^{-1}$ with the major axis pointing towards the Galactic center.
To simulate the RAVE data, we run a simulation for all the directions 
and then a subsample of stars is randomly chosen in each magnitude bin in the I filter, containing exactly the same number of stars observed by RAVE. 
A full comparison of this new Besan\c{c}on Galaxy Model with RAVE will be presented in a future paper (Robin et al. 2015, in prep.).\\ 

In Figure \ref{Density}, we present a comparison of the radial velocity relative to the Galactic rest-frame ($V_{GRV}$) distribution 
for the simulated sample and the RAVE data. To obtain $V_{GRV}$,
which is simply the star's radial velocity as seen from the Sun's location but in a reference frame at rest with the Galaxy,
we use the formulation given by \citet[][]{Hawkins+2015} and the following equation:
\begin{equation}
\label{GRV}
V_{GRV}\equiv RV + [U_{\odot}\cos(l) + (V_{\odot} + V_{LSR})\sin(l)]\cos(b) + W_{\odot}\sin(b)
\end{equation}

The Besan\c{c}on Galaxy Model reproduces very well the general kinematics observed in the RAVE data, 
as evidenced by the signal of the Disk observed in the sinusoidal-like density path seen in Figure \ref{Density}, 
with a higher density around $l=270^{\circ}$ and $V_{GRV} \sim -220$ km/s \citep[e.g.,][]{Hawkins+2015}.\\

Assuming a Gaussian velocity distributions and asymmetric drifts for the Thin-Disk, Young/Old Thick-Disk and Halo components, 
stars with $V_{total}> 180$ km s$^{-1}$ have a high probability of belonging to the Halo \citep[see][and references therein]{Nissen+2010}. 
However, in the case of the Besan\c{c}on Galaxy model \citep[][]{Robin+2014}, 
the velocity distribution of the Thick-Disk is non-Gaussian with an extended tail toward $V_{total}< 239 $ kms$^{-1}$, 
then in our simulated RAVE stars, those with  $180$ kms$^{-1} <V_{total} < $ 239 kms$^{-1}$ might belong to the Thick-Disk population. 
In this sense, since our selected stars have $V_{LSR}<-239$ km s$^{-1}$, they are most probably free of Thick-Disk contamination
(and certainly free of the Thin-Disk one), nonetheless we could expect Halo stars with preferential retrograde motion
in the Galactic longitude $240^{\circ} < l < 360^{\circ}$ explored in this work.\\

\section{Orbits in an axisymmetric Galactic model}
\label{experiment}

\begin{figure}
	
	\begin{center}
		\includegraphics[width=0.5\textwidth]{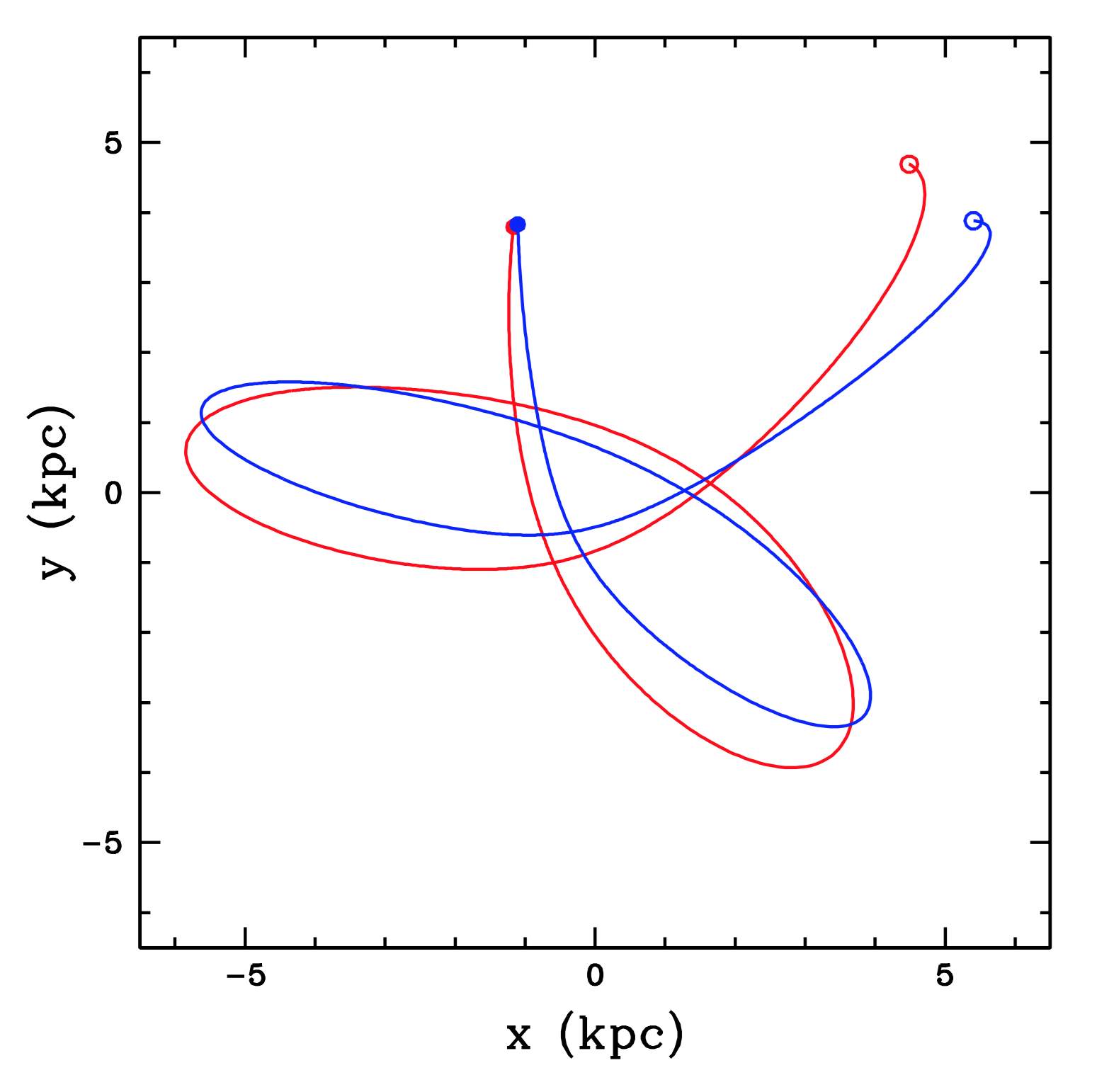}
	\end{center}
	\caption{Orbit projection in the potential axisymmetric for "J131340.4$-$484714" (blue line) and $\omega$ Centauri (red line) in X, Y Galactic coordinates on the Galactic plane. The solar position is (X$_{\odot}$, Y$_{\odot}$, Z$_{\odot}$) = (8.3, 0, 0) kpc in $t=0$ Gyr. The solid curves show the orbits integrated backwards from initial position in $t=0$ Gyr (open symbols) to $\tau_{enc}=0.2$ Gyr (filled symbols) having a close encounter with $d_{min}< 100$ pc. (A color version of this figure is available in the online journal.)} 
	\label{New3}
\end{figure}

In this work, we ran $10^5$ pairs of orbits for each RAVE star and $\omega$ Centauri cluster pair, using the Monte Carlo scheme\footnote{\textit{Uncertainties are considered as 1$\sigma$ variations, and a Gaussian Monte Carlo sampling generates the parameters to compute the present-day positions and velocities of the $\omega$ Centauri and RAVE stars}} describe in \citet[][]{Pichardo+2012}. For each Monte Carlo set of orbits pairs, the proper motions, radial velocities and distances, with their respective error bars, are used for both the star and the cluster. As initial test, we calculate orbits backward in time over 1 Gyr from the current position and velocity of $\omega$ Centauri in a Milky-Way-like axisymmetric Galactic potential (dynamical friction, spiral arms and bar effects are ignored), which is the  \citet{Allen+1991} model \citep[see details in][]{Pichardo+2012}, 
scaled with the new values $R_{0}=8.3$ kpc, $\Theta_{0}=239$ km$s^{-1}$ given by \citet{Brunthaler+2011}. Each Monte Carlo run employs a different scaled Galactic potential, using the uncertainties in $R_{0}$ and $\Theta_{0}$.
Then, we calculated the probability\footnote{The probability is defined as: $Prob =$ Number of orbits having close encounters with the cluster / $N_{total}$, where $N_{total} = 1\times10^{5}$ Monte Carlo simulations}
of close encounters in the past, which was defined as: (\textit{i}) having a minimum approach distance,  $d_{min} < 100 $ pc \citep[e.g.,][]{Fernandez-Trincado2015}, such that during the encounter the RAVE star is within the region of bounded stellar motions of $\omega$ Centauri, (\textit{ii}) having an encounter time, $\tau_{enc}< 0.2$ Gyr; in which the bar shape of the Galactic bulge and its effect on the Galactic orbits may be approximately ignored, this assumption is supported by the results in Appendix \ref{Append} (We verify that the introduction of a bar in our potential, using the prolate bar model of \citet{Pichardo+2004}, does not change significantly our conclusions for eight of our candidates with relative velocities $V_{rel}<200$ kms$^{-1}$ listed in Table \ref{table2}). $\omega$ Centauri has an orbital period $\tau_{orb}\sim0.08 - 0.12$ Gyr \citep[e.g.,][]{Dinescu+1999, Mizutani+2003} thus, in 0.2 Gyr the cluster will have approximately two perigalactic points. To illustrate the orbital behavior in about 0.2 Gyr, Figure \ref{New3} shows an example of a pair of orbits star-cluster for a RAVE star having a close encounter ($d_{min}< 100$ pc) with $\omega$ Centauri in $\tau_{enc} \sim 0.2$ Gyr and a relative velocity of $V_{rel}=53.6$ kms$^{-1}$. The average distance star-cluster is, $d_{avg} = 0.9$ kpc, from  $t=0$ Gyr until the close encounter time ($t=\tau_{enc}$).\\
		

For $\omega$ Centauri itself, we calculate the orbit, based on the current distance of $5.57\pm0.08$ kpc \citep[][]{Navarrete+2015}, 
proper motion $\mu_{\alpha}cos(\delta)=-5.08\pm0.35$ mas yr$^{-1}$ and $\mu_{\delta}=-3.57\pm0.34$ mas yr$^{-1}$, 
radial velocity 232.2$\pm$0.7 km s$^{-1}$ \citep{Dinescu+1999},  
and position $\alpha=$13h26m47.24s and $\delta=-47^\circ 28'46.5''$ \citep{Harris+1996}. \\

We briefly highlight some limitations of our calculation, since we ignore the effects of: 
$(i)$ dynamical evolution of the cluster, and tidal effects on the cluster exerted by the Galactic Disk in each orbit; 
$(ii)$ mass loss, as $\omega$ Centauri was likely more massive in the past \citep[see ][]{Bekki+2003, Ideta+2004}, this reduces the probabilities we find. 
We have also ignored secular changes in the Milky Way potential over time \citep{Lind+2015}.\\

\section{Results and discussion}
\label{discusion}

Within the sample of 329 stars with $V_{LSR}<-239$ kms$^{-1}$, 
a group of 15 RAVE stars were identified as having a probable encounter with $\omega$ Centauri within 0.2 Gyr at 100 pc distance, and another group of 314 stars as having unprobable ($<0.01\%$) close encounter with $\omega$ Centauri with the same distance and time limit.\\

 For those 15 stars with probable encounters, we identified 7 stars with high relative velocities $V_{rel}>200$ kms$^{-1}$, for which it is unlikely that they have been ejected from $\omega$ Centauri; and 8 stars were identified with relative velocities $V_{rel}<200$ kms$^{-1}$, which are more likely associated with $\omega$ Centauri (both groups of stars are listed Table \ref{table2}). These stars (8 more likely) have an average Galactic rotation velocity of $V_{LSR}\sim$ -280.9 $\pm$ 37.5 kms$^{-1}$, 
which is close to that of the $\omega$ Centauri, $V_{LSR}= -265.3 \pm 55.0$ kms$^{-1}$, as computed by us from the cluster's data.
The resulting values $U_{LSR}, V_{LSR}, $ and $W_{LSR}$ for these stars are given in Table \ref{table1}. 
Table \ref{table1} gives the same parameters for the 314 stars classified as 
unlikely to have been ejected from $\omega$ Centauri.\\

	
\begin{sidewaystable}
	\begin{tiny}
		\label{table1}
		\caption{$\omega-$Centauri candidates selected from RAVE catalogue and their abundance ratios, and space velocities. The last column indicate the number of Monte Carlo samples with close encounters within 0.2 Gyr at less than 100 pc with a relative velocity less than 200 kms$^{-1}$ at an axisymmetric Galactic potential.}
		\begin{tabular}{lllccccccccccccccc}
			\hline
			\hline
			RAVEID           &  $\alpha_{J2000}$      &  $\delta_{J2000} $   &  T$_{eff}$ &  log $g$ &  [Al/H]    &  [Si/H]   &  [Fe/H]    &  [Ti/H]    &  [Ni/H]    &  [Mg/H]    &  $U_{LSR}$      &  $V_{LSR}$      & $W_{LSR}$  & $D$ & $RV$ $^{\dag}$&  \textit{flag} \\
			&         hh:mm:ss            &      dd:mm:ss                &    (K)         &    dex     & dex        &    dex    &     dex      &  dex      &   dex      & dex  & km$^{-1} $& km$^{-1}$ & km$^{-1} $ & kpc & km$^{-1}$ &   \\   
			\hline
			\hline
			J131340.4$-$484714 & 13:13:40.43 & -48:47:13.8 &  4516.3 & 0.83 &  -     &  -0.75 &  -1.07 &  -1.02  & -     &  -0.77 &  13.3 $\pm$ 48.3 &  -279.2 $\pm$  41.5 &  -25.2 $\pm$ 48.1 & 4.4$\pm$ 1.1& 220.2 $\pm$ 1.1 &3,582 \\
			J120407.6$-$073227 & 12:04:07.64 & -07:32:26.8 &  4481.2 & 1.32 &  -0.43 &  -0.7  &  -0.71 &  -0.29  & -     &  -0.52 &  -34.5$\pm$ 28.4 &  -305.7 $\pm$  53.4 &  -15.4 $\pm$ 37.8 & 2.8$\pm$ 0.6& 160.1 $\pm$ 0.9 &1,266 \\
			J151658.3$-$123519 & 15:16:58.34 & -12:35:19.3 &  4481.3 & 1.41 &  -0.63 &  -0.81 &  -1.27 &  -0.6   & -     &  -0.64 &  50.3 $\pm$ 25.8 &  -246.0 $\pm$  74.2 &  0.02   $\pm$ 31.2 & 3.4$\pm$ 0.9& 64.9  $\pm$ 0.9 &1,059 \\
			J153016.8$-$214028 & 15:30:16.81 & -21:40:28.0 &  4544.8 & 1.05 &  -     &  -1.15 &  -1.33 &  -1.59  & -     &  -1.38 &  -26.9$\pm$ 28.8 &  -252.1 $\pm$  51.0 &  191.7 $\pm$ 30.1 & 3.2$\pm$ 0.7& 113.7 $\pm$ 0.5 &1,009 \\
			J143500.9$-$264337 & 14:35:00.85 & -26:43:36.8 &  4313.8 & 0.62 &  -1.0  &  -1.11 &  -1.21 &  -0.62  & -     &  -1.04 &  15.6 $\pm$ 21.5 &  -268.3 $\pm$  44.6 &  3.4   $\pm$ 22.9 & 4.9$\pm$ 0.8& 122.5 $\pm$ 0.9  &966  \\
			J150703.5$-$112235 & 15:07:03.45 & -11:22:34.8 &  4945.3 & 1.79 &  -     &  -0.7  &  -1.07 &  -0.97  & -     &  -0.45 &  -40.9$\pm$ 17.3 &  -241.1 $\pm$  59.9 &  -17.0 $\pm$ 18.2 & 2.1$\pm$ 0.5& -13.7 $\pm$ 2.4 &923  \\
			J040133.8$-$832428 & 04:01:33.83 & -83:24:27.9 &  4357.3 & 1.19 &  -0.33 &  -0.47 &  -0.9  &  -0.59  & -0.81 &  -0.59 &  -27.3$\pm$ 26.2 &  -243.5 $\pm$  32.1 &  65.4  $\pm$ 32.4 & 2.6$\pm$ 0.5& 148.6 $\pm$ 0.7 &902  \\
			J143024.9$-$085046 & 14:30:24.93 & -08:50:45.6 &  4629.0 & 1.56 &  -0.58 &  -0.63 &  -0.91 &  -0.69  & -     &  -0.39 &  -41.8$\pm$ 20.6 &  -334.8 $\pm$ 100.4 & -35.8  $\pm$ 18.6 & 1.7$\pm$ 0.5& 15.6  $\pm$ 0.8 &801  \\
			J144926.4$-$211745 & 14:49:26.41 & -21:17:44.7 &  4495.0 & 1.12 &  -0.86 &  -0.81 &  -0.99 &  -0.5   & -     &  -0.96 &  -15.5$\pm$ 24.4 &  -377.8 $\pm$  89.9 &  -48.9 $\pm$ 34.1 & 2.6$\pm$ 0.6& 78.1  $\pm$ 0.8 &727  \\
			J144618.8$-$711740 & 14:46:18.81 & -71:17:40.2 &  4384.9 & 1.27 &  -     &  -0.75 &  -1.22 &  -0.86  & -     &  -1.13 &  -91.6$\pm$ 50.2 &  -318.9 $\pm$  49.9 &  65.9  $\pm$ 28.6 & 3.1$\pm$ 0.7& 163.9 $\pm$ 0.8 &707  \\
			J152554.5$-$032741 & 15:25:54.50 & -03:27:41.0 &  4437.5 & 1.32 &  -     &  -0.64 &  -1.09 &  -0.69  & -     &  -0.48 &  12.1 $\pm$ 27.6 &  -264.6 $\pm$  79.0 &  81.2  $\pm$ 30.8 & 3.5$\pm$ 0.9& 51.4  $\pm$ 0.9 &643  \\
			J124722.9$-$282260 & 12:47:22.91 & -28:22:59.8 &  4754.8 & 2.0  &  -1.27 &  -1.09 &  -1.56 &  -1.65  & -     &  -1.22 &  2.4  $\pm$ 33.1 &  -295.1 $\pm$  39.7 &  59.9  $\pm$ 24.6 & 1.2$\pm$ 0.3& 241.2 $\pm$ 0.6 &635  \\
			J144734.5$-$722018 & 14:47:34.53 & -72:20:17.7 &  4285.1 & 0.58 &  -0.79 &  -0.48 &  -0.81 &  -0.46  & -0.4  &  -0.6  &  51.3 $\pm$ 28.4 &  -264.4 $\pm$  26.2 &  -27.5 $\pm$ 12.5 & 2.3$\pm$ 0.5& 235.9 $\pm$ 1.1  &577  \\
			J123113.7$-$194441 & 12:31:13.69 & -19:44:41.2 &  4609.3 & 1.93 &  -0.64 &  -0.5  &  -0.79 &  -0.31  & -     &  -0.34 &  -59.6$\pm$ 36.7 &  -257.9 $\pm$  47.1 &  12.9  $\pm$ 29.0 & 2.1$\pm$ 0.6& 158.0 $\pm$ 1.1 &569  \\
			J110842.1$-$715260 & 11:08:42.12 & -71:52:59.9 &  4386.0 & 0.62 &  -0.95 &  -     &  -1.33 &  -      & -0.9  &  -     &  44.7 $\pm$ 25.9 &  -263.5 $\pm$  12.1 &  -64.1 $\pm$ 21.7 & 2.7$\pm$ 0.6& 272.7 $\pm$ 0.8 &560  \\
			\hline
			\hline
		\end{tabular}  
		\tablefoot{$^{\dag}$ Heliocentric radial velocity ($RV$). Table \ref{table1} is published in its entirety in the electronic edition of the journal. A portion is shown here for guidance regarding its content.}
	\end{tiny}
	\end{sidewaystable}

We identified 314 RAVE stars with unlikely (or null) encounters with $\omega$ Centauri occurred within 100 pc. Figure \ref{Density} shows that their Galactocentric radial velocity (red symbols) is in agreement with the kinematics predicted by the Besan\c{c}on Galaxy model for the Halo. Their [Fe/H] and $V_{LSR}$ are in agreement with characteristics of the inner Galactic halo \citep[e.g.,][]{Dinescu+1999}.\\

Figure \ref{Density} shows stellar debris candidates (blue symbols) from \citet{Majewski+2012},
with a kinematical and chemical pattern linked to $\omega$ Centauri.
In the same figure, stars in our kinematical and chemical RAVE selection (orange symbols)
have retrograde orbits similar to that of $\omega$ Centauri, 
and are kinematically coherent with \citet{Majewski+2012} sample.
It is also evident that our 15 RAVE stars candidates with retrograde velocities 
do not follow the Disk kinematics in both Figures \ref{Toomre} and \ref{Density}.\\ 

We also identified in our sample one star with RAVEID$=$J131340.4$-$484714, 
previously identified beyond the cluster tidal radius by \citet{Anguiano+2015} and 
listed as potential member of the $\omega$ Centauri globular cluster. 
Under our numerical simulation scheme, we found 3,582/100,000 orbits for J131340.4$-$484714 having close encounters with $\omega$ Centauri, i.e. 
our results suggest that the star has been ejected from $\omega$ Centauri with an ejection velocity ($V_{ejection}\sim80$ kms$^{-1}$). Figure \ref{New1} shows that this star could have been ejected approximately $\sim$2.0 Myr ago, following an orbit very close to $\omega$ Centauri as is evident in the Figure \ref{New2} . This peculiar star could also be a rare "no-escaper" i.e. a star on a stable orbit outside the cluster tidal radius \citep{Ross1997}.\\

\begin{figure*}
	\begin{center}
		\includegraphics[width=1.0\textwidth]{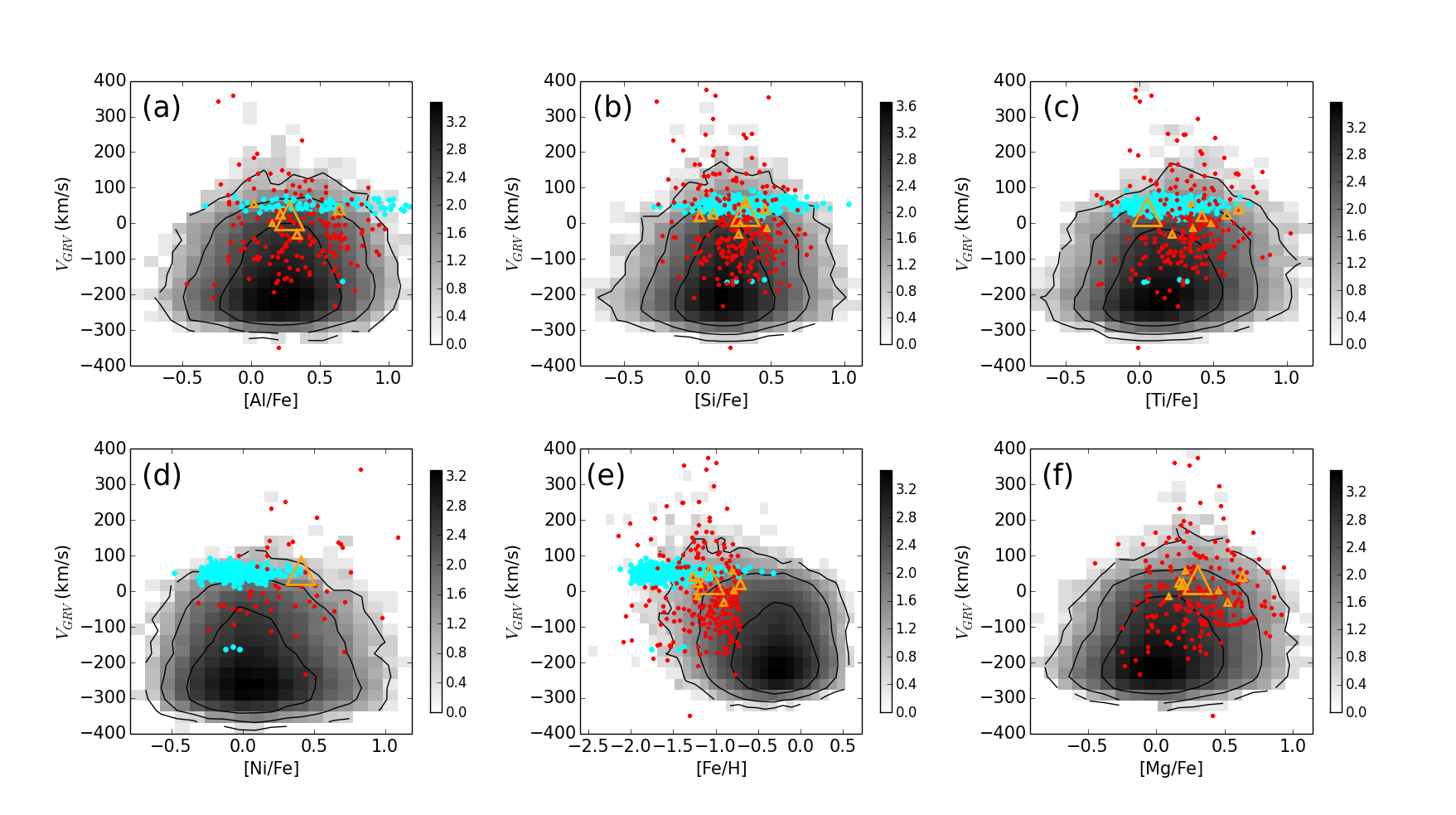}
	\end{center}
	\caption{2D density plot for  [X/H] and $V_{GRV}$, in the Galactic longitude range $240^{\circ} < l < 360^{\circ}$. 
		The orange and red symbols are the same as in Figure \ref{Toomre}. 
		The cyan dots refers $\omega$ Centauri members from \citet{Johnson+2010}. (A color version of this figure is available in the online journal.)}
	\label{Figure4}
\end{figure*}

\subsection{Close encounters: Star-Cluster}
\label{closeenco}

\begin{figure*}
	\begin{center}
		\includegraphics[width=1.1\textwidth]{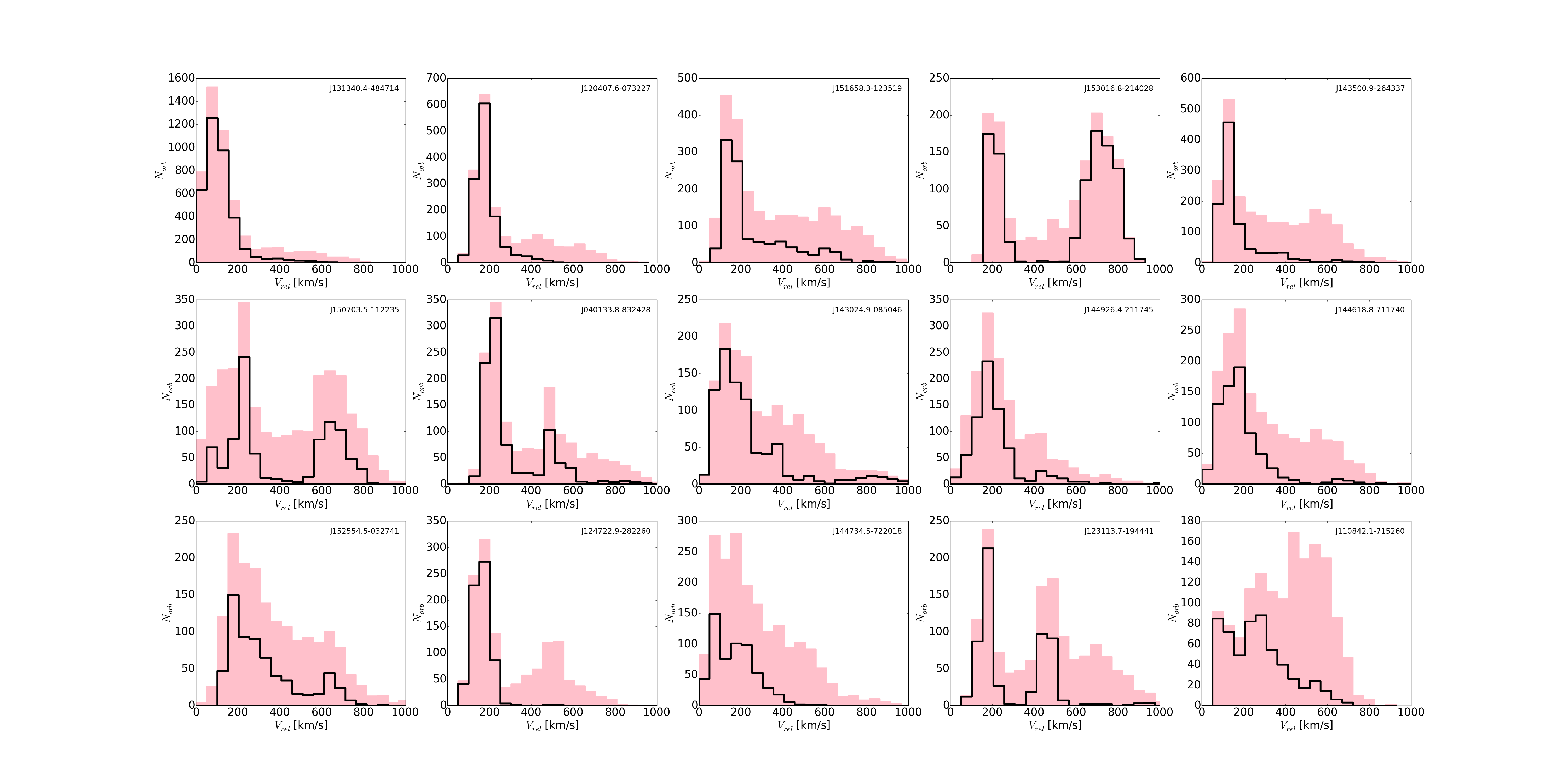}
	\end{center}
	\caption{Distribution of relative velocities $(V_{rel})$ during close encounters for the whole integration time, $t=1$ Gyr (pink histograms), and $t<0.2$ Gyr (black histograms). (A color version of this figure is available in the online journal.)} 
	\label{New1}
\end{figure*}

\begin{figure*}
	\begin{center}
		\includegraphics[width=1.1\textwidth]{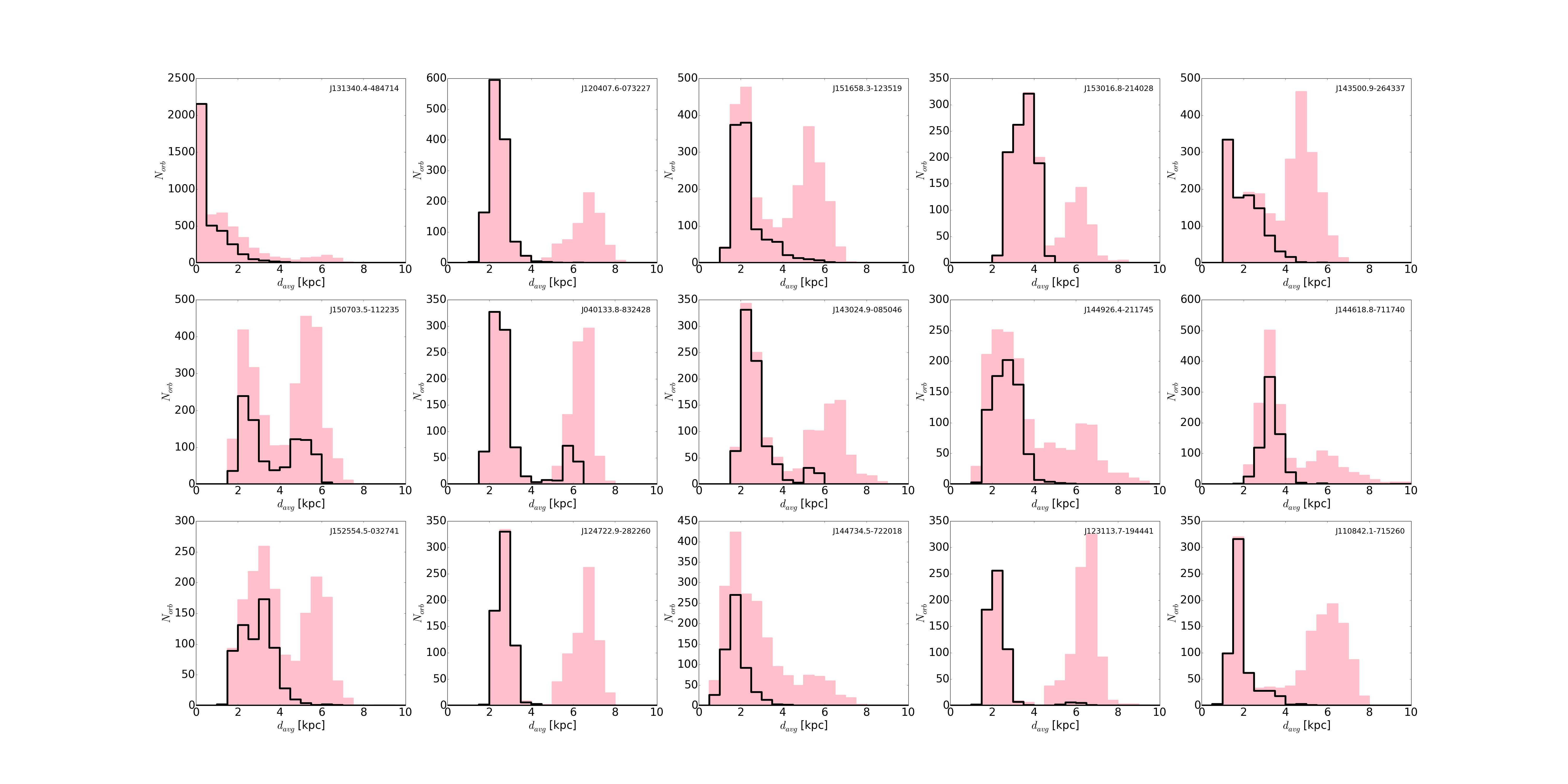}
	\end{center}
	\caption{$d_{avg}$ distributions, defined as the average distance between pairs of orbits through the Galaxy from the present-time (or $t=0$ Gyr) until the close encounter time ($\tau_{enc}$). The pink distribution refers to the whole integration time (1 Gyr), the black distribution refer to close encounters with relative velocities within 0.2 Gyr and 100 pc distance. (A color version of this figure is available in the online journal)} 
	\label{New2}
\end{figure*}

 Figures \ref{New1}, and \ref{New2} shows the relative velocity ($V_{rel}$) distribution for 15 RAVE stars having close encounters in the past, and the average distance ($d_{avg}$) distribution for the pairs of orbits from $t=0$ Gyr to $t=\tau_{enc}$, respectively. As is evident in Figure \ref{New1}, the vast majority of the close encounters are produced during approximately time intervals less than two times the orbital period of $\omega$ Centauri, with relative velocities $V_{rel}<200$ kms$^{-1}$ \citep[e.g.,][]{Lind+2015} in most cases, and closed orbits with small values of $d_{avg}$, such that the pairs of orbits (star - cluster) travel with close distances between them (see ilustrated example in Figure \ref{New3}), i.e., this approach clearly suggests that their orbits are 
 confined to the $\omega$ Centauri orbit. Table \ref{table2} lists 15 RAVE stars with the more frequent relative velocity ($V_{rel}$), average distance ($d_{avg}$), and 
 enconter time ($\tau_{enc}$) for each Monte Carlo simulation. \\

The escape velocity of $\omega$ Centauri cluster at the center is about $V_{esc,0}=60.4$ km s$^{-1}$, 
and at the cluster half-mass radius $V_{esc,h}=44$ km s$^{-1}$ \citep[e.g.,][]{Gnedin+2002}. 
Given the observational errors, and relative velocity $V_{rel} > V_{esc,0}$ star-cluster encounters, 
it is seems unlikely that the star has been tidally stripped from the cluster,
as this mechanism is not efficient to explain such fast encounters in our simulations.
For $V_{rel} > V_{esc,0}$, the star most possibly has been ejected from the cluster,
which is explained in some scenarios by binary system interactions and black holes \citep[e.g.][see references therein]{Pichardo+2012, Lind+2015}.  \\

\subsection{Chemical distribution for probable $\omega$ Centauri members }

Figure \ref{Figure4} shows $V_{GRV}$ vs. [Al/Fe], [Si/Fe], [Ti/Fe], [Ni/Fe], [Fe/H] and [Mg/Fe],
for the 15 RAVE stars (orange symbols) and the whole RAVE sample (in grey bins), 
The distribution for the 15 stars found as most likely debris from the $\omega$ Centauri progenitor,
clearly follows the $\omega$ Centauri chemical patterns,  
as observed for member stars within $\omega$ Centauri's tidal radius \citep[e.g.,][and references therein]{Johnson+2010}. 
The heavy elements (Si and Ti) generally follow the typical enhancement of $\sim$ +0.3 dex found in the cluster population.  
$\omega$ Centauri exhibits a complex history of chemical enrichment, 
with an abundant metal poor and intermediate population \citep[e.g,][]{Johnson+2010}. 
The extended enhancement in chemical abundances observed in our sample of 329 RAVE stars
is not conclusive to select $\omega$ Centauri members from RAVE data,
however, we have shown on the Toomre diagram of Figure \ref{Toomre}, that these stars follow the Halo kinematics. 
 It should be noted that $\omega$ Centauri spans ranges in all the analysed abundance patterns
which are similar to those of the Milky Way, so any similarity between these quantities is not (poorly)
useful to confirm the origin of these stars.

\begin{table*}
	\setlength{\tabcolsep}{4.0mm}  
	\begin{tiny}
		\caption{Derived ejection parameters for the 15 RAVE stars more likely having close encounters with $\omega$ Centauri.}
		\begin{tabular}{ccccccc}
			\hline
			\hline
			RAVEID &  & Axisymmetric Potential & & & Non-axisymmetric Potential & \\
			\hline
			\hline
                            &   $V_{rel}$  &  $d_{avg}$  & $\tau_{enc}$ &   $V_{rel}$  &  $d_{avg}$  & $\tau_{enc}$\\  
             & kms$^{-1}$ & kpc & Myr & kms$^{-1}$ & kpc & Myr\\
     \hline        
 	\hline
 	Encounters with $V_{rel}<200$ kms$^{-1}$ &  &  & & &  & \\
 	\hline
      J131340.4$-$484714 &   75.9  &  0.25  &   2.4     & 79.7   & 0.26 & 2.1\\
      J120407.6$-$073227 &   170.3 &  2.25  &  37.4   & 143.2   & 2.28 &  36.1\\
      J151658.3$-$123519 &   137.9 &  2.3   &  35.4    & 139.5   & 1.88 &  32.1\\
      J143500.9$-$264337 &   108.3 &  1.32  &  71.4   & 108.2   & 1.35 &  71.1\\
      J143024.9$-$085046 &   105.6 &  2.34  &  28.4   &  182.7  & 2.34 &  28.1\\
      J144618.8$-$711740 &   166.8 &  2.89  &  88.4   &  132.9  & 2.87 &  97.1\\
      J144734.5$-$722018 &     77.0  &  1.34  &  45.4  &  187.6  & 1.81 &  43.1\\
      J123113.7$-$194441 &   167.8 &  2.31  &  30.4   &  170.4  &  2.31&  31.1\\
      \hline
      \hline
      Encounters with $V_{rel}>200$ kms$^{-1}$\\
       \hline
      J150703.5$-$112235 &   224.7 &  2.36  &  23.4 &  227.9 & 2.39 &   24.1 \\
      J040133.8$-$832428 &   225.2 &  2.34  &  29.4 &  237.7 & 2.34 &   29.1 \\
      J144926.4$-$211745 &   200.8 &  2.34  &  39.4 &  229.9 & 2.79 &   39.0 \\      
      J153016.8$-$214028 &   678.0 &  2.84  & 118.4&  718.6 & 3.48 & 122.1 \\
      J152554.5$-$032741 &   206.0 &  2.84  &  48.4 &  323.7 & 4.02 &   57.1 \\
      J124722.9$-$282260 &   207.3 &  2.38  &  46.4 &  208.2 & 2.85 &   48.1 \\
      J110842.1$-$715260 &   275.0 &  1.34  &  45.4 &  291.5 & 1.91 &   45.1  \\
\hline
\hline
\end{tabular}  \label{table2}
\end{tiny}
\end{table*}

\subsection{Statistical significance}

	To establish the statistical significance of our results we compare the Monte Carlo simulations (see \S\ref{experiment}) to five mock samples created using the Besan\c{c}on Galaxy model \citep{Robin+2014}. For each mock sample we chose 329 stars randomly satisfying the selection criteria defined in \S\ref{selection}, and the RAVE/UCAC4 selection bias (errors in radial velocities, proper motions and distances are added). The number of false positive are listed in Table\ref{tablefinal}.\\
	
\begin{table}
	\setlength{\tabcolsep}{4.0mm}  
	\begin{tiny}
		\caption{Mock sample from Besan\c{c}on Galaxy model \citep{Robin+2014}.}
		\begin{tabular}{cc}
			\hline
			\hline
			Id  & Number of false positives ($N_{false}$) \\
			\hline
			\hline
		   Mock sample 1  & 4 stars \\
		   Mock sample 2  & 2 stars \\
		   Mock sample 3  & 2 stars \\
		   Mock sample 4  & 4 stars \\
		   Mock sample 5  & 1 stars \\
			\hline
			\hline
		\end{tabular}  \label{tablefinal}
	\end{tiny}
\end{table}

We consider a high significance in our detection if 

\begin{equation}
\sigma_{significance} = N_{true} - \langle{}N_{false}\rangle{}   > 3\times{}\sigma_{false}
\end{equation}

\noindent
where,  $N_{true}= 15$ (see text), and as we have 5 mock samples from Table \ref{tablefinal}, we take the average over all the samples ($\langle{}N_{false}\rangle{} = 2.6 $), obtaining a mean and standard deviation ($\sigma_{false} = 1.2$) for this values.  Then, we find a significance $\sigma_{significance} = 10.33$. Therefore, the number of stars reported in the present paper are $\sigma_{significance} > 3\times{}\sigma_{false}$ compared to the mock sample. We can conlude that the methodology in \S\ref{experiment} robustly show that there is a statistically significant population of stars from RAVE survey possibly ejected from $\omega$ Centauri.\\

\section{Conclusion}
\label{conclusion}

We identified 8 RAVE stars at galactic longitudes ($240^{\circ} < l < 360^{\circ}$), 
with [Fe/H] metallicities in the range -2.2 dex $<$[Fe/H]$< -0.7$ dex, 
with Galactic retrograde motion similar to that of $\omega$ Centauri,
 which could have had a close encounter with $\omega$ Centauri 
with a minimum approach $d_{min}< 100$ pc during the last 200 Myr in the past.
Our numerical scheme suggest that these stars have been probably ejected from the cluster with a velocity $V_{ejection}< 200$ kms$^{-1}$, and their kinematics connection with $\omega$ Centauri makes them good candidates to be part of the predicted sequence of tidal streams in retrograde rotation left behind the cluster over time.\\

The probabilities we estimated may neglect some physical evolution of the orbit and/or internal structure of the $\omega$ Centauri progenitor,
but the agreement with other results in the literature supports our conclusion that the 8 RAVE stars reported in this study
are candidates to be $\omega$ Centauri stellar debris.\\

 We have also identified another group of 7 stars having close encounters with $\omega$ Centauri, however the high relative velocities $V_{rel}> 200 $ kms$^{-1}$ during the encounters, makes it unlikely that they have been ejected from $\omega$ Centauri.\\

The other 314 stars with retrograde orbits have a negligible probability of close encounters within 100 pc in 0.2 Gyr. 
We have classified this group of stars as less likely to have ties with $\omega$ Centauri.
Their chemical distribution and Galactic retrograde rotation are similar to those of the inner Galactic halo.\\

We have not found an evident external structure beyond the tidal radius of the cluster from RAVE data on Galactic retrograde rotation.\\

  In order to facilitate the reproducibility and reuse of our results, 
  we have made available all the orbit simulations in a public repository\footnote{\url{http://fernandez-trincado.github.io/Fernandez-Trincado/simulations.html}}.\\
  
\begin{acknowledgements} 

J.G.F-T is currently supported by Centre National d'Etudes Spatiales (CNES) through Ph.D grant 0101973 and the R\'egion de Franche-Comt\'e, and by the French Programme National de Cosmologie et Galaxies (PNCG). This research was supported by the Munich Institute for Astro- and Particle Physics (MIAPP) of the DFG cluster of excellence "Origin and Structure of the Universe". J.G.F-T also thanks to James Binney and George Kordopatis for discussions of this work during the "THE NEW MILKY WAY" workshop. F. Robles-Valdez is grateful to DGAPA-UNAM for the postdoctoral grant CJIC/CTIC/0903 and acknowledge the support from the CONACyT (grant 167625).
We are grateful to an anonymous referee for a careful reading and many useful comments that significantly improved the final manuscript.
Besan\c{c}on Galaxy Model simulations were executed on computers from the UTINAM Institute of the Universit\'e de Franche-Comte, 
supported by the R\'egion de Franche-Comt\'e and Institut des Sciences de l'Univers (INSU). 
Monte Carlo simulations were executed on computers from the Instituto de Astronom\'ia-UNAM, M\'exico. Funding for RAVE has been provided by: 
the Australian Astronomical Observatory,
the Leibniz-Institut fuer Astrophysik Potsdam (AIP),
the Australian National University,  
the Australian Research Council, 
the French National Research Agency,
the German Research Foundation (SPP 1177 and SFB 881), 
the European Research Council (ERC-StG 240271 Galactica),
the Istituto Nazionale di Astrofisica at Padova, 
the Johns Hopkins University, 
the National Science Foundation of the USA (AST-0908326), 
the W. M. Keck foundation,
the Macquarie University,
the Netherlands Research School for Astronomy,
the Natural Sciences and Engineering Research Council of Canada,
the Slovenian Research Agency,
the Swiss National Science Foundation, 
the Science \& Technology Facilities Council of the UK,
Opticon, Strasbourg Observatory and 
the Universities of  Groningen, Heidelberg and Sydney. 
The RAVE web site is at \url{http://www.rave-survey.org}\\ 
\end{acknowledgements}

\bibliographystyle{aa}
\bibliography{references.bib}


\begin{appendix}
	
\section{Orbits in a non-axisymmetric Galactic model}	
\label{Append}	
	
\begin{figure*}
		\begin{center}
	\includegraphics[width=1.1\textwidth]{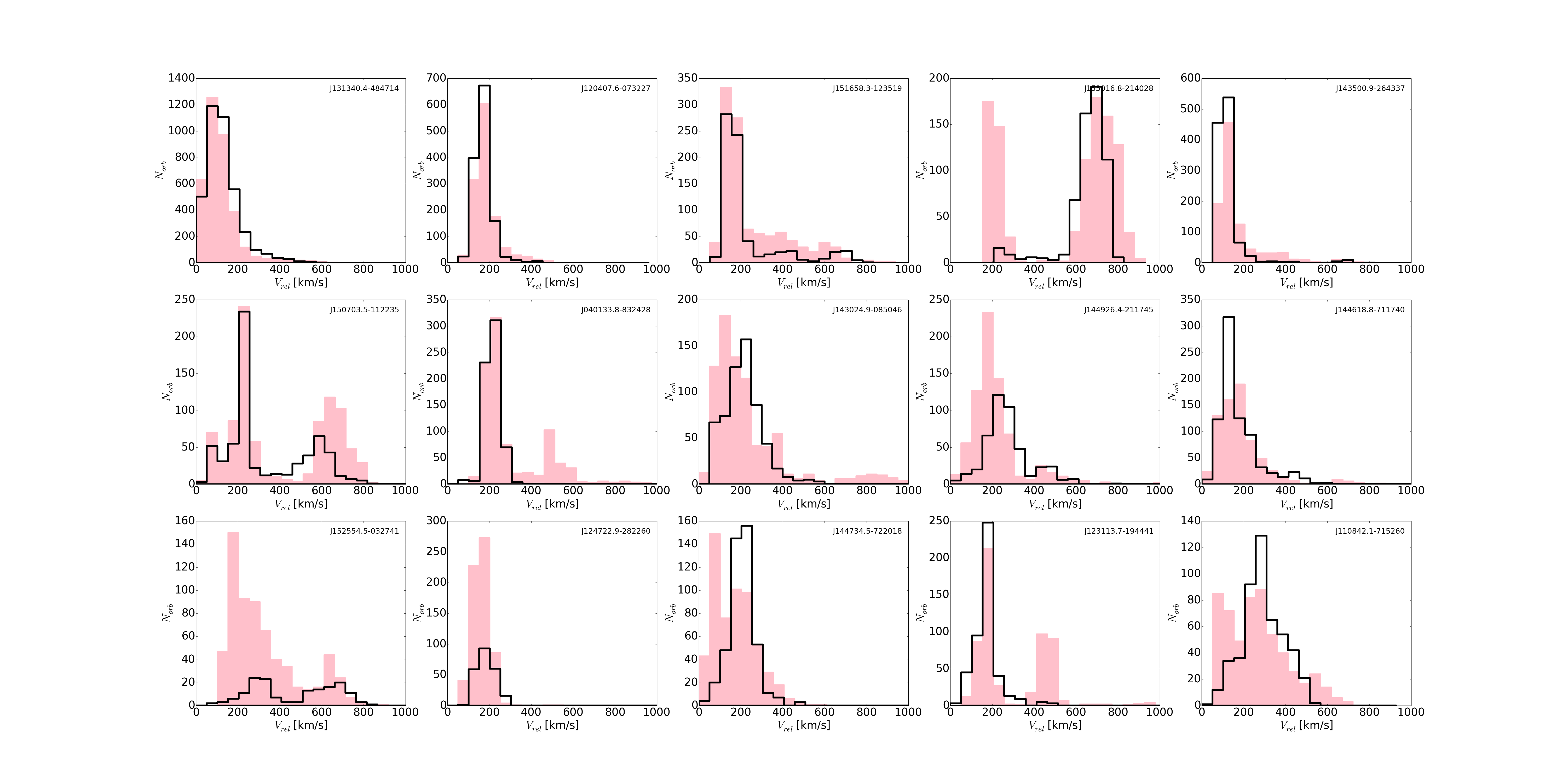}
\end{center}
\caption{Relative velocity ($V_{rel}$) distribution for 10$^{5}$ pairs of orbits integrated backward in time over 0.2 Gyr in an axisymmetric (pink histograms) and non-axisymmetric potential (black histograms).} 
\label{nonaxisymmetric1}
\end{figure*}

\begin{figure*}
		\begin{center}
			\includegraphics[width=1.1\textwidth]{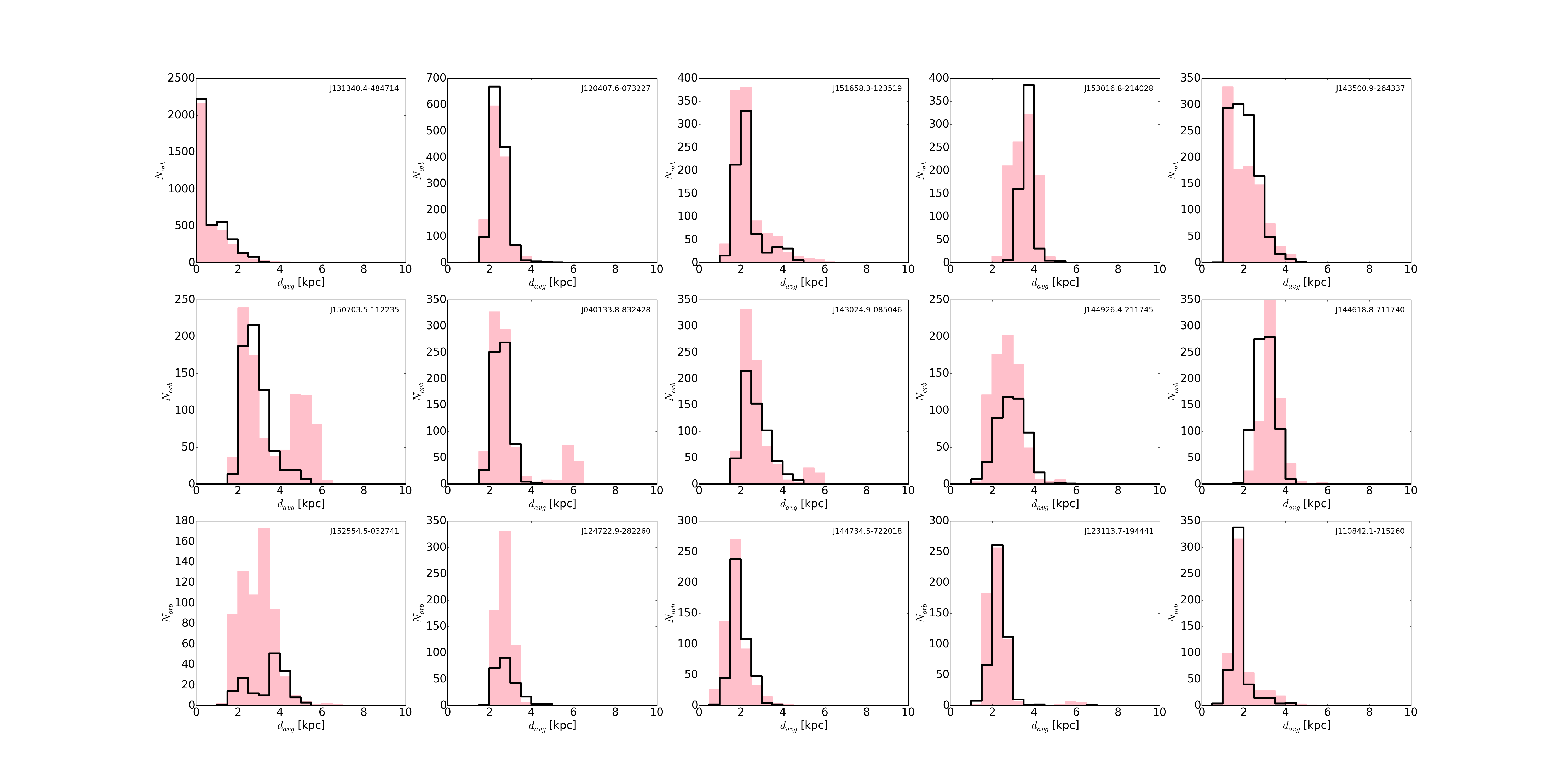}
		\end{center}
		\caption{ $d_{avg}$ distribution. The pink and red histograms are the same as in Figure \ref{nonaxisymmetric1}.} 
		\label{nonaxisymmetric2}
	\end{figure*}

We also run the simulations in a non-axisymmetric galactic potential, 
in order to verify the bar effect in our results. The Galactic model that we 
considered is the analytical potential presented in \S\ref{experiment}, and the non-axisymmetric Galactic potential including a prolate bar, given by \citet{Pichardo+2004} to the galactic bar, with a total mass of $\sim$1.6$\times10^{10}$ M$_{\odot}$ and angular velocity of $\Omega_{B}=55\pm5$ kms$^{-1}$kpc$^{-1}$ \citep[e.g.,][]{Pichardo+2012}. Using the same considerations that \S\ref{experiment} to model the orbits of the RAVE stars and $\omega$ Centauri, 10$^{5}$ pairs of orbits were integrated to 0.2 Gyr in the past. Figure \ref{nonaxisymmetric1} and Figure \ref{nonaxisymmetric2} shows the results to the stars listed in Table \ref{table2}; the probabilities of close encounters are similar in both Galactic potential, and the bar effects for integration time $<0.2$ Gyr, does not affect the results presented in this work.  

\end{appendix}	

\end{document}